\documentstyle[12pt]{article}

\makeatletter
\@addtoreset{equation}{section}

\makeatother

\def\beq{\begin{equation}}
\def\eeq{\end{equation}}
\def\bea{\begin{eqnarray}}

\def\eea{\end{eqnarray}}
\def\ds{\displaystyle}

\def\ssz{\scriptsize}

\def\req#1{(\ref{#1})}
\def\refs#1{sect. \ref{#1}}

\def\ie{ i.e. }
\def\eg{ e.g. }
\def\rep{{\rm Re}\ }

\def\Tr{{\rm Tr}\ }

\begin{document}

\begin{center}
{\Large \bf Zeta-Regularization of the $O(N)$ Non-Linear Sigma
Model in $D$ dimensions}   \\
running title: $O(N)$ Non-Linear Sigma Model in $D$ dimensions \\
\hskip1cm   \\
{\bf Emili Elizalde$^{a,b}$},
{\bf Sergei D. Odintsov$^{b},$}\footnote{On
 leave of absence from Tomsk Pedagogical Institute,
634041 Tomsk, Russia. E-mail: sergei@ecm.ub.es}
and {\bf August Romeo$^a$}
\\
$^a$Center for Advanced Studies, CEAB, CSIC,
Cam\'{\i} de Santa B\`arbara, 17300 Blanes \\
$^b$Department ECM and IFAE,
Faculty of Physics, University of  Barcelona, \\
Diagonal 647, 08028 Barcelona, Catalonia, Spain
\end{center}

\begin{abstract}
The O(N) non-linear sigma model in a $D$-dimensional space of the form
${\bf R}^{D-M} \times {\bf T}^M$,
${\bf R}^{D-M} \times {\bf S}^M$,
or ${\bf T}^M \times {\bf S}^P$
is studied, where  ${\bf R}^M$, ${\bf T}^M$ and ${\bf S}^M$
correspond to flat space, a torus and a sphere, respectively. Using
zeta regularization and the $1/N$ expansion, the corresponding partition
functions ---for deriving the free energy---
and the gap equations are obtained.
In particular, the free energy at the criticial point on
${\bf R}^{2q+1} \times {\bf S}^{2p+2}$ vanishes in accordance
with the conformal equivalence to the flat space ${\bf R}^D$.
Numerical solutions of the
gap equations at the critical coupling constants are given, for several
values of $D$. The properties of the partition function and its
asymptotic behaviour for large $D$ are discussed. In a similar way, a
higher-derivative non-linear sigma model is investigated too.
The physical relevance of our results is discussed.
\end{abstract}

\begin{center} PACS 02.30.+g, 02.30.Mv, 03.65.-w \end{center}

\newpage

\section{Introduction}

Zeta regularization \cite{1a} (for a review see \cite{2a}) is a
very powerful and elegant method for regularizing the divergences that
appear in quantum field theory. It has found lots of different
applications, from the calculation of the vacuum energy density or
Casimir energy corresponding to very different configurations (different
fields, different spacetimes, different boundaries) to its application
in wetting/nonwetting phenomena in actual condensed matter and solid
state systems, to the analysis of phase transitions coming from the
study of effective potentials in different context, topological mass
generation, Bose-Einstein condensation phenomena, evaluation of the
partition function in string and p-brane theories, etc. \cite{2a}.
{}From a more
mathematical point of view, the method has allowed the computation of
the basic operator tr log $(\Box + X)$ on a curved manifold, that is
very important in quantum gravity and cosmology.

In this paper we will study the zeta-function regularization of
the $O(N)$ non-linear sigma model in an arbitrary number, $D$, of
dimensions and on spaces of the form
${\bf R}^{D-M} \times {\bf T}^M$,
${\bf R}^{D-M} \times {\bf S}^M$,
and ${\bf T}^M \times {\bf S}^P$.
In flat spacetime this model has a big number of different applications
(see \cite{3a} for a review), in particular, in the theory of critical
phenomena and solid state physics.
For example, such a three-dimensional model, which is known to be
renormalizable in this case, may be used in condensed matter physics
as an effective field theory of the two-dimensional quantum
antiferromagnet \cite{Fr}. In particular, the $O(N)$ non-linear sigma
model on ${\bf S}^1 \times {\bf R}^2$ may be applied to describe
the low-temperature properties of the quantum antiferromagnet \cite{NePe}.
Recently, in connection with the
study of higher-dimensional conformal theories \cite{Ca} ---where one
can use well-known $2d$ conformal field theory techniques--- such a model
has been considered in three dimensional curved spacetime \cite{GRV}. The
critical properties of the model were also  discussed.

Here we will extend this important analysis, by studying the $O(N)$
non-linear sigma model in topologically non-trivial spaces of constant
curvature in arbitrary dimension $D$. Using zeta-function regularization
and the $1/N$ expansion techniques we will obtain the partition function
and the gap equation in each of the cases considered, and also numerical
solutions of the gap equations for some spaces. The asymptotic behaviour
in the limit $D\rightarrow \infty$ will be considered too. Finally, a
higher-derivative generalization of the $O(N)$ non-linear sigma model will
be introduced and the corresponding zeta-function on a flat but
topologically non-trivial space will be obtained. Numerical solutions of
the gap equations in this last model, for different values of $D$, will
be discussed too.
The relevant partition function and gap equations are presented in
sects. 2 and 3, while the applications to
${\bf R}^{D-M} \times {\bf T}^M$ and
${\bf R}^{D-M} \times {\bf S}^M$ appear in sects. 4 and 5,
respectively.  Our higher-derivative model is introduced in sect. 6.
In the conclusion we comment on the relevance of our results and
discuss future perspectives.
The appendix is devoted to the calculation of the necessary zeta
functions for the spaces under consideration.

\section{The $O(N)$ nonlinear sigma model in $D$ dimensions: \
\ \ partition function}

Let us consider an arbitrary $D$-dimensional space  of constant (or
zero) curvature. It is well-known that the scalar conformally
invariant
D'Alembertian operator on such manifold is
\beq -\Box+\xi R, \eeq
where $\xi={ D-2 \over 4(D-1)}$. In what follows we use Euclidean
space
notations.
The partition function of the $O(N)$ nonlinear sigma model in $D$
dimensions we will be interested in is given as follows:
\beq
Z[g] = \int {\cal D}\phi {\cal D}\sigma \exp\left\{
-{1 \over 2 \lambda} \int d^Dx \sqrt{g} \left[
\phi^i ( -\Box+\xi R ) \phi^i + \sigma (\phi^{i \ 2}-1) \right]
\right\},
\label{Ztheor}
\eeq
where $\phi^i$ are scalars in curved spacetime, $i=1, \dots, N$,
$\xi={ D-2 \over 4(D-1)}$ is the conformal coupling, $\sigma$ is an
auxiliary scalar introduced in order to keep the constraint
$\phi^i(x) \phi^i(x)=1$, coming from the condition of
$O(N)$-invariance,
and $\lambda$ is the coupling constant. Note that $\sigma$ has no
dynamics, as it just plays the role of a Lagrange multiplier.
Observe also that we have choosen to work with the conformally
invariant
operator $-\Box+\xi R$, in order to better understand the
conformally invariant properties of the model.

The theory \req{Ztheor} was extensively studied in \cite{GRV} in
three-dimensional curved spacetime at its nontrivial fixed point.
In
particular it was shown, in the large-$N$ limit, that such a model
is
an example of a conformal field theory at a nontrivial fixed point.
Investigation of this
theory in different spaces of constant curvature has suggested that
what distinguishes a given model is not curvature, but the
conformal
class of its metric.

Our purpose here will be to study the theory in $D$-dimensional
curved
spacetimes of constant curvature near the nontrivial fixed point in
the
$1/N$-expansion. What is even more important, we will analyze, in
addition
to the large-$N$ limit, some situations involving the limit $D
\to\infty$, which is relevant to dimensional dependence
investigations in a number of quantum systems \cite{Mi,Be}.
Numerical solutions to the gap equations will also be given.

It is convenient to rescale $\phi\to\sqrt{\lambda}\phi$.
Then
\beq
Z[g] = \int {\cal D}\phi {\cal D}\sigma \exp\left\{
-\int d^Dx \sqrt{g} \left[
{1 \over 2} \phi^i ( -\Box+\xi R +\sigma ) \phi^i
-{\sigma  \over 2 \lambda}
\right] \right\} ,
\label{Zresc}
\eeq
where the mass-dimensions of the fields and parameters are
\beq
[ \phi ]={ D-2 \over 2}, \hspace{1cm} [ \sigma ]=2, \hspace{1cm}
\left[ 1 \over \lambda \right]=D-2, \hspace{1cm} [ \xi ]=0,
\eeq
and the dependence of $Z$ on the metric $g_{\mu \nu}$
is explicitly shown.
Note that sometimes it is convenient to rewrite the partition
function
\req{Zresc} as explicitly regularized. In particular, if one uses
a
cutoff $\Lambda$ for regularizing, it may be adequate to do the
change
$ {1 \over \lambda( \Lambda )} \to
{\Lambda^{D-2} \over \lambda( \Lambda )}$
in order to work with a dimensionless $\lambda( \Lambda )$.

The aim is to study the above theory in the large-$N$ limit
keeping, as usual, $N\lambda$ fixed as $N \to \infty$. The
spacetime
dimension $D$ will be arbitrary.
Integrating out the first $N-1$ components of $\phi$ and
rescaling the $N$th component $\phi_N$ to
$\sqrt{N-1 \over 2}\phi_N$,
and ${N-1 \over 2}\lambda$ to $\lambda$, we get
\beq
\begin{array}{ll}
\ds Z[g]=&\ds \int {\cal D}\phi_N {\cal D}\sigma \
\exp \left( -{N-1 \over 2} \left\{
{\rm Tr} \log ( -\Box+\xi R +\sigma )  \right. \right. \\
&\ds \left. \left. +{1 \over 2}\int d^Dx \sqrt{g} \left[
\phi_N ( -\Box+\xi R +\sigma ) \phi_N
-{\sigma  \over 2 \lambda}
\right] \right\} \right) .
\label{action}
\end{array}
\eeq

\section{The gap equations}
Since we shall deal with manifolds of constant curvature, we will
look
for a uniform saddle point: $\sigma(x)=m^2$, $\phi_N(x)=b$.
Extremizing the action \req{action}
with respect to $\phi_N(x)$ maintaining
$\sigma(x)$ fixed, and the other way around, we obtain the
{\it gap equations}
\beq
\begin{array}{rcl}
( -\Box+\xi R+m^2 )b&=&0, \\
\ds G(x,x; m^2, g)+b^2-{1 \over \lambda}&=&0,
\end{array}
\eeq
where
\beq
G(x,x; m^2, g)= \langle x| ( -\Box+\xi R+m^2 )^{-1} |x \rangle
\eeq
is the two-point Green function at equal points. Once the solutions
to
these equations have been found, it is sensible to evaluate the
free
energy density $W$ at the saddle point to leading order in $1/N$
\beq
W[g, \lambda]=
{N \over 2}\left[ \Tr\log ( -\Box+\xi R+m^2 )
-\int d^Dx \sqrt{g} {m^2 \over \lambda} \right] .
\eeq

Applying zeta-function regularization one defines
\beq
\begin{array}{rcl}
\ds G(x,x; m^2, g)=\lim_{s \to 1}G(x,x;m^2,g;s)&\propto&
\ds\lim_{s \to 1}\langle x| \zeta_{\cal M}(s) |x \rangle , \\
\Tr\log ( -\Box+\xi R+m^2 )&\propto&-\zeta_{\cal M}'(0).
\end{array}
\label{GTrLzr}
\eeq
$\zeta_{\cal M}(s)$ is the spectral zeta function of the operator
$-\Box+\xi R+m^2$ on the spacetime ${\cal M}$ under consideration,
\ie
\beq
\begin{array}{rcl}
\ds\zeta_{\cal M}(s)&=&\Tr( -\Box+\xi R+m^2 )^{-s} .
\end{array}
\eeq
The proportionality factors ---not explicitly written in
\req{GTrLzr}---
are determined, in each case, by the normalization of the physical
states, and they have such a form that the dimensionalities match.

By heat-kernel series analysis, it is known that the short-distance
divergences of this two-point Green function depend in general on
the
curvature of the spacetime,
except for the leading pole which is independent of $g$ and,
therefore,
present in all cases.
As a result, in
flat spacetimes with nontrivial topology such as ${\bf T}^D$ or
${\bf R}^{D-M} \times {\bf T}^M$, this singular behaviour will be
the
same as in the flat space ${\bf R}^D$. That is why it is natural to
study ${\bf R}^D$ first.
According to this reasoning, the critical value of $\lambda$ ---at
which
the theory becomes finite--- will be the same for all these
spacetimes.

The gap equations in the spacetime ${\bf R}^D$, after momentum
cutoff
regularization, read
\beq
\begin{array}{rcl}
\ds m^2 b&=&0, \\
\ds b^2&=&\ds
\left[ {\Lambda \over \lambda(\Lambda)}-G_{\Lambda}(x,x; m^2, g)
\right].
\end{array}
\eeq
Here $G_{\Lambda}(x,x; m^2, g)$ means the Green function obtained
when setting a cutoff $\Lambda$ on the norm of the integrated
momentum. Studying their solutions, one finds that, for $b=m=0$,
\beq
{\Lambda \over \lambda_c(\Lambda)}=
\int^{(\Lambda)} {d^Dk \over (2 \pi)^D} {1 \over k^2}
={1 \over (4 \pi)^{D/2} \Gamma\left( D \over 2 \right) }
{ \Lambda^{D-2} \over {D\over 2}-1 } , \ \mbox{if $D > 2$} ,
\eeq
and, for $b=0$, $m \neq 0$,
\beq
{\Lambda \over \lambda_c(\Lambda)}-{\Lambda \over \lambda(\Lambda)}
=m^2 \int^{(\Lambda)} {d^Dk \over (2 \pi)^D} {1 \over k^2(k^2+m^2)}
=-{m^{D-2} \over (4 \pi)^{D/2}} \Gamma\left( 1-{D \over 2} \right)
+\epsilon (\Lambda),
\eeq
where $\epsilon (\Lambda)$ are
terms vanishing as $\Lambda\to\infty$.
Since $\Gamma\left( 1-{D \over 2} \right) < 0$ for odd $D>1$, this
indicates the unphysical character of the solution when
$\lambda < \lambda_c$. A more detailed study shows that
$\lambda = \lambda_c$ is a critical value separating two different
phases.

In the same spacetime, when using zeta-function regularization
the second gap equation becomes
\beq
\begin{array}{rcl}
\ds b^2&=&\ds\lim_{s \to 1}
\left[ {1 \over \lambda(s)}-G(x,x; m^2, g; s) \right],
\end{array}
\eeq
where
\beq
G(x,x; m^2, g; s)=\int {d^Dk \over (2\pi)^D} {1 \over (k^2+m^2)^s}
={m^{D-2s} \over (4\pi)^{D \over 2}}
{ \Gamma\left( s-{D \over 2} \right) \over \Gamma(s) } .
\eeq
Following the line of thinking of ref \cite{GRV},
we set the values $m=b=0$ from the discussion in cutoff
regularization, and realize that now the only consistent way out is
\beq \lim_{s \to 1}{1 \over \lambda_c(s)}=0 , \eeq
which gives the critical value of $\lambda$ in this regularization.

In curved spacetimes without boundaries,
heat-kernel expansion gives
\beq
\begin{array}{c}
\ds G_{\Lambda}(x,x;m^2,g)=
\int_{1/\tilde\Lambda^2}^{\infty} dt \
\langle x| e^{-t( -\Box+\xi R + m^2 )} | x \rangle = \\
\left\{
\begin{array}{ll}
\ds{1 \over (4 \pi)^{D/2}}\left[
{a_0 \over {D\over 2}-1} \tilde\Lambda^{D-2}
+{a_2 \over {D\over 2}-2} \tilde\Lambda^{D-4}
+ \dots
+{a_{D-1} \over {1\over 2}} \tilde\Lambda^{2}
\right]
+ f_1 (\Lambda ),
&\mbox{for odd $D$}, \\
\ds{1 \over (4 \pi)^{D/2}}\left[
{a_0 \over {D\over 2}-1} \tilde\Lambda^{D-2}
+{a_2 \over {D\over 2}-2} \tilde\Lambda^{D-4}
+ \dots
+a_{D-4} \tilde\Lambda^2
+a_{D-2} \log \tilde\Lambda^2
\right]
+ f_2 (\Lambda ),
&\mbox{for even $D$,}
\end{array}
\right.
\end{array}
\label{Ghk}
\eeq
where $f_1(\Lambda) $ and $f_2 (\Lambda )$ are terms that become
finite
when $\Lambda\to\infty$ and where, for consistency, we have taken
$\tilde\Lambda^2
= \left[ \Gamma\left( D \over 2 \right) \right]^{-{D-2 \over 2}}
\Lambda^2$.
As usual, the $a_{2n}$'s stand for the even
Seeley-Gilkey coefficients \cite{SG},
and we have taken into account that $a_{2n+1}=0$ in the absence of
boundaries. Since $a_0=1$, for the first $D$'s we immediately get
\begin{center}
\begin{tabular}{l|l}
$D$&divergences of $G_{\Lambda}(x,x;m^2,g)$ \\ \hline
3&${1 \over (4\pi)^{3/2}} 2\tilde\Lambda $ \\
4&${1 \over (4\pi)^2} \left( \tilde\Lambda^2+a_2\log
\tilde\Lambda^2 \right) $ \\
5&${1 \over (4\pi)^{5/2}}
    \left( {2\over 3}\tilde\Lambda^3 +2a_2\tilde\Lambda \right) $
\\
6&${1 \over (4\pi)^3}
    \left( {1 \over
2}\tilde\Lambda^3+a_2\tilde\Lambda^2+a_4\log\tilde\Lambda^2
\right)$ \\
7&${1 \over (4\pi)^{7/2}}
    \left( {2\over 5}\tilde\Lambda^5 +{2 \over 3}a_2\tilde\Lambda^3
           +2a_4\tilde\Lambda \right) $ \\
\end{tabular}
\end{center}
The $D=3$ result does not depend on $a_2, a_4, \dots$ and is thus
independent of the curvature. In consequence, it is the same for
any
3-dimensional space without boundaries and therefore it is enough
to
find it in ${\bf R}^3$.
For conformally flat manifolds of any $D$, the
property $R=0$ makes $a_2,a_4,\dots$ vanish with the same
consequence,
\ie, we can get by with the critical values in ${\bf R}^D$.
This fact will be used in the study of ${\bf R}^{D-M}\times {\bf
T}^M$.

When $R\neq 0$ in $D>3$, a particular study of every particular
situation is called for.
In spaces of constant positive curvature, such as spheres
or some products of ${\bf R}^D$ by spheres,
the first gap equation reads $(\xi R+m^2)b=0$ with $R>0$, and
admits
no other solution than $b=0$. From the second equation,
$\ds{\Lambda \over \lambda( \Lambda )}=G_{\Lambda}(x,x;m^2,g)$,
where
the divergent parts of the r.h.s. as $\Lambda\to\infty$ are given
by
\req{Ghk}. Next, we form the difference between the $m=0$ and the
$m > 0$ cases:
\beq
{\Lambda \over \lambda_{m=0}( \Lambda )}
-{\Lambda \over \lambda( \Lambda )}
={1 \over (4 \pi)^{D/2}}\left[
{a_2(m=0)-a_2(m) \over {D\over 2}-2} \tilde\Lambda^{D-4}
+{a_4(m=0)-a_4(m) \over {D\over 2}-4} \tilde\Lambda^{D-6}
+ \dots \right] .
\eeq
To obtain $a_{2n}(m)$ from $a_{2n}(m=0)$ is just a matter of
replacing
$\xi \to \xi+{m^2 \over R}$ in every expression of these
coefficients.
As an example we consider $D=5$, whose divergences appear in the
preceding table. Since $a_2(m=0)=\left( \xi -{1 \over 6 } \right)
R$,
$a_2(m=0)-a_2(m)=m^2$ independently of the value of $R$. After
dividing
by $\Lambda$ we get
\beq
{1 \over \lambda_{m=0}( \Lambda )}
-{1 \over \lambda( \Lambda )}
={1 \over (4 \pi)^{5/2}} {\pi^{3/4} \over 8} 2 m^2 .
\eeq
This equality stops making sense if $\lambda < \lambda_{m=0}$,
which
can be interpreted by regarding $\lambda_{m=0}=\lambda_c$ as the
critical value separating two phases. Since this was obtained by
setting $b=0$ and $m=0$, those are the values we shall set for
computing ${1 \over \lambda_c}$ in zeta regularization \cite{GRV}.

\section{${\bf R}^{D-M}\times {\bf T}^M$ \label{sectRT}}
Putting the adequate normalization factors, the Green function and
the
free energy at the critical value of $\lambda$ read
\beq
\begin{array}{rcl}
\ds G(x,x;m^2,g;s=1)&=&\ds{1 \over (2\pi)^{D-M} \rho^M}
\zeta_{{\bf R}^{D-M}\times {\bf T}^M}(1), \\
\ds {W \over N}&=&\ds -{1 \over 2}\left( \rho \over 2\pi
\right)^{D-M}
{\zeta_{{\bf R}^{D-M}\times {\bf T}^M}}'(0).
\end{array}
\label{GWzs}
\eeq
After calculating the zeta function for our operator on this
spacetime
(see the Appendix), we can write
\beq
\zeta_{{\bf R}^{D-M}\times{\bf T}^M}(s)=
{\pi^{D/2} \over \Gamma(s)}
\left( 2 \pi \over \rho \right)^{-2s+D-M} \left[
I_M\left( s-{D-M \over 2}, {\rho m \over 2\pi} \right)
+ \left( \rho m \over 2\pi \right)^{D-2s}
\Gamma\left( s-{D \over 2} \right)
\right] .
\label{zIMGa}
\eeq
The second term tells us that the singularities at $s=0,1$ are
present
only when $D$ is even, independently of $M$.
$I_M\left( s-{D-M \over 2}, {\rho m \over 2\pi} \right)$, is an
integral of the type
\beq
I_M(z, \alpha)=
\int_0^{\infty} dt \ t^{z-M/2-1} \ e^{-t \alpha^2}
\left[ \theta^M\left( \pi \over t \right) -1 \right] ,
\label{defIM}
\eeq
where
$\theta(x)$ is the Jacobi function
\beq
\theta(x)=\sum_{n=-\infty}^{\infty} e^{-\pi x n^2} .
\label{deftheta}
\eeq
The
key point is that the integrand is well-behaved around $t=0$,
causing no new pole at $z=M/2$ (\ie at $s=D/2$).
Further, $I_M$ can be expanded into a Dirichlet series, which
reads
\beq
I_M\left( s-{D-M \over 2}, {\rho m \over 2\pi} \right)=
2^{D/2-s} \left( \rho m \over 2\pi \right)^{D-2s}
\sum_{l=1}^M \left( \begin{array}{c} M \\ l \end{array} \right)
2^{l+1} \sum_{ \vec{n} \in ({\bf M}^*)^l }
{ K_{D/2-s}( \rho m |\vec{n}|_l ) \over
( \rho m |\vec{n}|_l  )^{D/2-s} } ,
\label{IMDir}
\eeq
where $K$ is the modified Bessel function
and $| \vec n |_l$ stands for the Euclidean norm of
$\vec{n}=( n_1, \dots, n_l )$.

The particular cases we will calculate are those of odd $D=2d+3$,
$M=1$.
Under these conditions the Bessel functions involved are just
\[
K_{{D \over 2}-1}(x)= K_{d+{1 \over 2}}(x)=
\sqrt{ \pi \over 2x } e^{-x}
\sum_{k=0}^d { (d+k)! \over k! (d-k)! } {1 \over (2x)^k} ,
\]
and we may interchange the summations
to obtain
\beq
\begin{array}{c}
\ds\zeta_{{\bf R}^{2d+2}\times{\bf T}^1}(1)=
\pi^{d+3/2}
\left( 2 \pi \over \rho \right)^{2d}
\left( \rho m \over 2\pi \right)^{2d+1} \\
\ds\times\left[
2^{d+2} \sqrt{\pi}
\sum_{k=0}^d { (d+k)! \over k! (d-k)! 2^k}
{ {\rm Li}_{d+1+k}(e^{-\rho m}) \over ( \rho m )^{d+1+k} }
+\Gamma\left( -d-{1 \over 2} \right)
\right] , \\
\ds{\zeta_{{\bf R}^{2d+2}\times{\bf T}^1}}'(0)=
\pi^{d+3/2}
\left( 2 \pi \over \rho \right)^{2d+2}
\left( \rho m \over 2\pi \right)^{2d+3} \\
\ds\times\left[
2^{d+3} \sqrt{\pi}
\sum_{k=0}^{d+1} { (d+1+k)! \over k! (d+1-k)! 2^k}
{ {\rm Li}_{d+2+k}(e^{-\rho m}) \over ( \rho m )^{d+2+k} }
+\Gamma\left( -d-{3 \over 2} \right)
\right] ,
\end{array}
\label{zetas01rtodd1}
\eeq
where ${\rm Li}$ is the polylogarithm function.
Setting $d=0$ ($D=3$), these formulas reproduce
---as should be expected--- the ones in \cite{GRV}.

We look at the solutions of the gap equations for the critical
value
${1 \over \lambda_c}=0$. The chances are:
\begin{enumerate}
\item{$m=0, b\neq 0$.}
This leads to
\[ b^2=-{\pi^{d+3/2} \over (2 \pi)^{2d+3}} {1 \over \rho^{2d+1}}
4\sqrt{\pi} {(2d)! \over d!} \zeta(2d+1). \]
When $d=0$ the r.h.s. diverges and no solution can exist. For $d
>0$,
$\zeta(2d+1)>0$ and there is a sign conflict which prevents the
appearance of any solution on this side.
\item{$m>0, b=0$.} Now we are posed with solving
\beq
\sum_{k=0}^d { (d+k)! \over k! (d-k)! 2^k}
( \rho m )^{d-k} {\rm Li}_{d+1+k}(e^{-\rho m})
+( \rho m )^{2d+1}
{ \Gamma\left( -d-{1 \over 2} \right) \over 2^{d+2} \sqrt{\pi} }
=0 .
\eeq
\end{enumerate}
Only $d=0$ admits a relatively simple analytic solution
because ${\rm Li}_1$ is the only polylogarithm which can be
trivially
expressed in terms of elementary functions. In that case one gets
(see also \cite{S,GRV}) $( \rho m)_c=2\log\tau$, with
$\tau= {1 + \sqrt{5} \over 2}$, \ie $( \rho m)_c \simeq 0.9624$.
Of course, it is also possible to find this same value by
numerically
solving
the above equation for $d=0$, and this is precisely what we do for
the next $d$'s. Afterwards, the value found is replaced into
\req{zetas01rtodd1} and \req{GWzs}, so as to find the free energy
$W$.
We thus arrive at:
\begin{center}
\begin{tabular}{r|r|r}
$d$&$(\rho m)_c$&${W \over N}(\rho m=(\rho m)_c)$ \\ \hline\hline
0&0.9624&$-0.1530$ \\ \hline
1&no solution& \\ \hline
2&2.1775&$-0.0441$ \\ \hline
3&no solution& \\ \hline
4&3.5504&$-0.0561$ \\ \hline
5&no solution& \\ \hline
6&3.6841&$-2.2634$ \\ \hline
\end{tabular}
\end{center}
The figure for ${W \over N}$ when $d=0$ coincides with the
numerical
value of $-{2 \over 5 \pi}\zeta(3)$, derived in
\cite{GRV} with the help of polylogarithm identities from
\cite{Lew}.
The rest of the values are the first (and possibly only)
solutions found after scanning a reasonable positive range.

We can now study the asymptotic behaviours of these expressions for
$M$  (and $D$) and/or $\alpha$ ($=\rho m/(2\pi)$) going to
infinity. Two cases will be considered: (i) $\alpha \gg 1$ and $M$
bounded, with $M-D$ fixed, and (ii) $\alpha \gg 1$ and $M \gg 1$ with
$\alpha /M \rightarrow $ const., again with $M-D$ finite.

Let us start with the first case. As everywhere the dependence on
$M$ is through $D-M$, it is enough to study $I_M (\tau, \alpha)$,
where $\tau =s-(D-M)/2$ with $s=0$ or $s=1$. From the behaviour
\beq
K_{D/2-s} (2\pi \alpha |\vec{n}|) \sim \left( \frac{\pi}{4 \pi
\alpha |\vec{n}|} \right)^{1/2} e^{-2\pi \alpha |\vec{n}|},
\eeq
we easily obtain that
\beq
I_M (\tau, \alpha) \sim \pi^{s-D/2} \alpha^{(D-1)/2-s} \sum_{l=1}^M
\left( \begin{array}{c} M \\ l \end{array} \right) 2^l
\ {\rm Li}_{(d+1)/2-s}^{(l)}(e^{-2\pi\alpha}),
\eeq
Where ${\rm Li}_{P}^{(l)}(x)$ denotes the generalized
polylogarithm, for which we have
\beq
{\rm Li}_P^{(l)}(x) =\sum_{n_1,\ldots,n_l=1}^\infty
\frac{x^{\sqrt{n_1^2+\cdots +n_l^2}}}{\sqrt{n_1^2+\cdots+n_l^2}^P}
\sim \frac{x^{\sqrt{l}}}{\sqrt{l}^P} + {\cal O}
\left(\frac{x^{\sqrt{l+3}}}{\sqrt{l+3}^P} \right), \ \ \ \ \ x \ll 1.
\eeq
Taking this into account, we obtain
\beq
I_M \left( s- \frac{D-M}{2}, \alpha\right) \leq 2 \pi^{s-D/2}
\alpha^{(D-1)/2-s} M^2 e^{-2\pi\alpha}.
\eeq
We can now consider $M$ to be large, of course, but never competing
with $\alpha$ or the above expression loses its sense.
In order to deal with both limits at the same time, we must
consider the case (ii). From the well-known behaviour of
$K_\nu (\nu z)$ for constant $z$ and $\nu \rightarrow \infty$,
by calling
\beq
u_l = \lim_{\alpha,D \rightarrow \infty} \frac{2\pi \alpha
\sqrt{l}}{D/2-s} \equiv \mbox{const.}, \ \ \ \ \ \
v_l=\eta(u_l)= \sqrt{1 +u_l^2} + \log \frac{u_l}{1+ \sqrt{1 +u_l^2}}
\equiv \mbox{const.},
\eeq
we get
\beq
I_M \left( s- \frac{D-M}{2}, \alpha\right) \sim
\frac{\sqrt{2}}{(1+u_1^2)^{1/4}} \left( \frac{u_1}{2} \right)^{D/2-
s}  \left(\frac{D}{2}-s\right)^{D/2-s+1} e^{-(D/2-s)v_1}.
\eeq

\section{${\bf R}^{D-M}\times {\bf S}^M$ \label{sectRS} }

The Green function and the  Tr log  contribution to the free
energy are now
\beq
\ds G(x,x;m^2,g;1)={1 \over (2\pi)^{D-M} a^M}
\zeta_{{\bf R}^{D-M}\times {\bf S}^M}(1),
\label{GFRD-MSM}
\eeq
and
\beq
\ds {W \over N} =
\ds -{1 \over 2}\left( a \over 2\pi \right)^{D-M}
{\zeta_{{\bf R}^{D-M}\times {\bf S}^M}}'(0),
\label{FRD-MSM}
\eeq
respectively, where $a$ is the radius of the sphere.
Here we will consider the massless case only.
By our discussion on gap equations, in $D=5$
$\ds {1 \over \lambda_c}=G(x,x;0,g,1)$. Therefore, in these
conditions,
$\zeta_{{\bf R}^{5-M}\times {\bf S}^M}(1)$ for $m=0$ tells us
the critical value of $\lambda$.
Another reason for the $m=0$ choice, apart from simplicity, is
that the $M=D-1$ case is conformally equivalent to ${\bf R}^D- \{
0 \}$,
and then, $m=0$ corresponds also to solutions for critical
$\lambda$.

Taking the conformal coupling and the known form of the Riemann
scalar
for the sphere, we follow
a method analogous to ref. \cite{CC} (see also \cite{ASD})
to construct
the required zeta function, which is
\beq
\zeta_{{\bf R}^{D-M} \times {\bf S}^M}(s)=
2 a^{2s-D+M} \pi^{D-M \over 2}
{ \Gamma\left( s-{D-M \over 2} \right) \over \Gamma(s)}
\times \left\{ \begin{array}{rll}
\ds \Sigma_{\alpha}(p, 2s-D+M), &\mbox{for $M=2p+2$}, \\
\ds \Sigma_{\beta}(p, 2s-D+M), &\mbox{for $M=2p+1$}, \\
\end{array} \right.
\eeq
where
\beq
\begin{array}{llrl}
\ds\Sigma_{\alpha}\left( p, 2z \right)&=&
\ds{1 \over (2p+1)!}&\ds\sum_{k=0}^{p}(-1)^k \alpha_k(p-1)
\zeta_H\left( -2p+2k+2z-1, {1 \over 2} \right), \\
\ds\Sigma_{\beta}\left( p, 2z \right)&=&
\ds{1 \over (2p)!}&\ds\sum_{k=0}^{p-1}(-1)^k \beta_k(p-2)
\zeta_R\left( -2p+2k+2z \right) ,
\end{array}
\label{DefSigmaalphabeta}
\eeq
$\zeta_H$ and $\zeta_R$ denote the Hurwitz and Riemann zeta
functions,
and the $\alpha_k$ and $\beta_k$ coefficients are
\beq
\begin{array}{llllllll}
\alpha_0(j)&=&1,&\hspace{1cm}&\alpha_k(j)&=&
\ds\sum_{0 \leq i_1 < \dots < i_k \leq j}
\left( i_1+{1 \over 2} \right) \cdots \left( i_k+{1 \over 2}
\right),
& \ k \geq 1, \\
\beta_0(j)&=&1,&\hspace{1cm}&\beta_k(j)&=&
\ds\sum_{0 \leq i_1 < \dots < i_k \leq j}
\left( i_1+1 \right) \cdots \left( i_k+1 \right),
& \ k \geq 1.
\end{array}
\label{defalphakbetak}
\eeq
Our notation is just slightly different from that in
ref. \cite{CC}. In fact
$\alpha_k(p-1)={1 \over 2^k}\alpha_k^{\rm CC}(p-1)$ and
$\beta_k(p-2)=\beta_k^{\rm CC}(p-1)$, where {\rm CC} stands for the
coefficients employed in \cite{CC}.
These coefficients enable one to write
\beq
d(M,l)=\sum_{k=0}^{k_{\mbox{\ssz max}}(M)} (-1)^k {\cal A}_k(M)
\left( l +{M-1 \over 2} \right)^{M-1-2k},
\label{dAks}
\eeq
where ${\cal A}_k(M)=2/(2p+1)! \, \alpha_k(p-1)$
and $k_{\mbox{\ssz max}}(M)=p$  for $M=2p+2$, while
${\cal A}_k(M)=2/(2p)! \, \beta_k(p-2)$
and $k_{\mbox{\ssz max}}(M)=p-1$
for $M=2p+1$.

In order to study the Green function and the  Tr log  contribution
to the free energy we need to evaluate the above zeta function at
$s=1$ and its derivative at $s=0$. Such quantities will be
expressed in
terms of
\beq
\begin{array}{llrl}
\ds{\Sigma_{\alpha}}' \left( p, 2z \right)&=&
\ds{1 \over (2p+1)!}&\ds\sum_{k=0}^{p}(-1)^k \alpha_k(p-1)
\zeta_H'\left( -2p+2k+2z-1, {1 \over 2} \right), \\
\ds{\Sigma_{\beta}}' \left( p, 2z \right)&=&
\ds{1 \over (2p)!}&\ds\sum_{k=0}^{p-1}(-1)^k \beta_k(p-2)
\zeta_R' \left( -2p+2k+2z \right) ,
\end{array}
\label{DefSigmaalphabetaprime}
\eeq
and the results will follow
Then, we have the following cases
\begin{enumerate}

\item{Even $D-M=2q$}

\begin{enumerate}

\item{$M=2p+2$}
\beq
\begin{array}{lll}
\ds\zeta_{{\bf R}^{2q}\times{\bf S}^{2p+2}}( 1+\varepsilon )&=&
\ds
\pi^q
{(-1)^{q-1} \over (q-1)!} 2 a^{-2q+2}
\left\{
\left[ {1 \over \varepsilon}+\gamma+\psi(q)+2\log{a     \mu}
\right]
\Sigma_{\alpha}(p,-2q+2) \right. \\
&&\ds\left.\hspace{8em}+ 2 \Sigma_{\alpha}'(p,-2q+2) \right\}
+O(\varepsilon),
\\
\ds{\zeta_{{\bf R}^{2q}\times {\bf S}^{2p+2}}}'(0)&=&
\ds
\pi^q
{(-1)^q \over q!} 2 a^{-2q}
\left\{
\left[ \gamma+\psi(q+1)+2\log{a     \mu} \right]
\Sigma_{\alpha}(p,-2q) \right. \\
&&\ds\left.\hspace{6em}+ 2 \Sigma_{\alpha}'(p,-2q) \right\} .
\end{array}
\eeq
As usual, $\mu$ is a parameter with mass dimension,
introduced by redefining
$\zeta_{\cal M}^{-\Box +\xi R}(s)$ as
$\mu^{-2s}\zeta_{\cal M}^{(-\Box +\xi R)/\mu^2}(s)$,
which renders the $\log$ arguments dimensionless.

\item{$M=2p+1$}
\beq
\begin{array}{lll}
\ds\zeta_{{\bf R}^{2q}\times {\bf S}^{2p+1}}(1)&=&
\ds
\pi^q
{(-1)^{q-1} \over (q-1)!}
2 a^{-2q+2}
\ 2 \Sigma_{\beta}'\left( p, -2q+2 \right) ,
\\
\ds{\zeta_{{\bf R}^{2q}\times {\bf S}^{2p+1}}}'(0)&=&
\ds
\pi^q
{(-1)^q \over q!}
2 a^{-2q}
\ 2 \Sigma_{\beta}'\left( p, -2q \right) .
\end{array}
\eeq
The absence of terms with primeless $\Sigma_{\beta}$
is a consequence of the location of the real zeros of
$\zeta_R$.

\end{enumerate}

\item{Odd $D-M=2q+1$}

\begin{enumerate}

\item{$M=2p+2$}
\beq
\begin{array}{lll}
\ds\zeta_{{\bf R}^{2q+1}\times {\bf S}^{2p+2}}(1)&=&0, \\
\ds{\zeta_{{\bf R}^{2q+1}\times {\bf S}^{2p+2}}}'(0)&=&0 .
\end{array}
\label{thisva}
\eeq
This vanishing follows from known properties of $\zeta_H(x,1/2)$  and,
as a result of \req{FRD-MSM},
\beq
\ds {W \over N}= 0.
\eeq

\item{$M=2p+1$}
\beq
\begin{array}{lll}
\ds\zeta_{{\bf R}^{2q+1}\times {\bf S}^{2p+1}}(1)&=&
\ds
\pi^{q+{1 \over 2}}
\Gamma\left( -q+{1 \over 2}\right)
{2 a^{-2q+1} \over (2p)!} \Sigma_{\beta}(p,-2q+1), \\
\ds{\zeta_{{\bf R}^{2q+1}\times {\bf S}^{2p+1}}}'(0)&=&
\ds
\pi^{q+{1 \over 2}}
\Gamma\left( -q-{1 \over 2} \right)
{2 a^{-2q-1} \over (2p)!} \Sigma_{\beta}(p,-2q-1) .
\end{array}
\eeq

\end{enumerate}

\end{enumerate}

When studying the four kinds of sums \req{DefSigmaalphabeta} and
\req{DefSigmaalphabetaprime} for
$p\to\infty$ and finite $2z$ of the type $-2q-1,-2q,-2q+1,-2q+2$,
one
has to consider
the behaviours of the $\alpha_k$ and $\beta_k$ coefficients, which
vary
within the ranges
\beq
\begin{array}{ccccrcl}
\alpha_0(p-1)&=&1,&\dots&,\alpha_p(p-1)&=&
\ds{\Gamma^2\left( p+ {1 \over 2} \right) \over \pi}, \\
\beta_0(p-2)&=&1,&\dots&,\beta_{p-1}(p-2)&=&\left[ (p-1)! \right]^2
,
\end{array}
\eeq
and also satisfy
\beq
\alpha_k(p-1) \sim \beta_k(p-2) \sim {p^{3k} \over 3^k k!}, \
\mbox{as $p\to\infty$}.
\eeq
For the $\beta_k$ coefficients, this property was already observed
in
\cite{CC}.

We must also take into account
the following asymptotics:
\beq
\begin{array}{lll}
\zeta_R(-(2n+1))&\sim&\ds (-1)^{n+1} {2 (2n+1)! \over
(2\pi)^{2n+1}}, \\
\zeta_H(-(2n+1),1/2)&\sim& -\zeta_R(-(2n+1)), \\
\zeta_R'(-(2n+1))&\sim& -\zeta_R(-(2n+1)) \log n, \\
\zeta_H'(-(2n+1),1/2)&\sim& -\zeta_R'(-(2n+1)) ,
\end{array}
\eeq
which are valid for $n \gg 1$, and follow from known results about
the gamma, Riemann and Hurwitz zeta functions.

We show, as an example, the case of $\Sigma_{\beta}(p, -2q+1)$.
Including the prefactor $1/(2p)!$, we denote the terms in that sum
by
$\ds
\Sigma_{\beta}(p, -2q+1)=(-1)^{p+q} \sum_{k=0}^{p-1} t_k,
$
where, as one may check for large $p$,
$ 0 < t_0< \dots < t_{p-1} $.
Therefore
\beq
\left\vert \Sigma_{\beta}(p, -2q+1) \right\vert < p \ t_{p-1},
\eeq
Combining the information we have with the Stirling
approximation for the factorial (or $\Gamma$) functions, we get
\beq
p \ t_{p-1}=
p \ {2 (2q+1)! \over (2 \pi)^{2q+2}! }\zeta(2q+2)
{ \left[ (p-1) \right]^2 \over (2p)! } \sim
{ (2q+1)! \zeta(2q+2)  \over 2^{2q+1} \pi^{2q+3/2} }
{p^{-1/2} \over 2^{2p}}.
\eeq
As a result,
\beq
\begin{array}{rcl}
\Sigma_{\beta}(p, -2q+1)&\to&0, \ \ \ \mbox{for any finite positive
$q$},  \\ p&\to&\infty .
\end{array}
\eeq
The other three types of sum have the same property but, since the
proof
is of similar nature, the details are omitted.
In consequence,
\beq
\begin{array}{rcl}
\ds\zeta_{{\bf R}^{D-M}\times{\bf T}^M}(1)&\to& 0, \\
\ds{\zeta_{{\bf R}^{D-M}\times{\bf T}^M}}'(0)&\to& 0, \\
M&\to&\infty, \ \mbox{finite ($D-M$)} ,
\end{array}
\eeq
\ie the two-point Green function at equal points and the $\Tr\log$
contribution to the free energy vanish in this case of the
$D\to\infty$ limit. The importance of that limit lies in the chance
of using the spacetime dimension as a perturbation parameter in
field theory, with the advantage that it is then possible to obtain
nonperturbative results in the coupling constants \cite{Be}, such
as Green functions for quantum fields in the Ising limit.
In this spirit, expansions in inverse powers of the
dimension have proven quite useful in atomic physics \cite{BMP}.

The $m>0$ case is mathematically more involved. A possible way out
is the construction of a power series in $am$ by combining the
preceding results with a simple binomial expansion. Such a method
leads to
\beq
\begin{array}{c}
\ds \zeta_{{\bf R}^{D-M} \times {\bf S}^M}(s)=
2 a^{2s-D+M} { \pi^{D-M \over 2} \over \Gamma(s) }
\sum_{k=0}^{\infty} {(-1)^k \over k!} (am)^{2k}
\Gamma\left( s+k-{D-M \over 2} \right) \\
\ds \times \left\{ \begin{array}{rll}
\ds \Sigma_{\alpha}(p, 2s+2k-D+M), &\mbox{for $M=2p+2$}, \\
\ds \Sigma_{\beta}(p, 2s+2k-D+M), &\mbox{for $M=2p+1$}. \\
\end{array} \right.
\end{array}
\eeq

\subsection{A calculation in ${\bf R}^2\times {\bf S}^3$}
For ${\bf S}^3$ the degeneracy of each spherical
mode (see \req{degsph}) is just $(l+1)^2$. After applying standard
Mellin-transform
techniques, we end up by writing the zeta function as follows:
\beq
\zeta_{{\bf R}^2\times {\bf S}^3}(s)=
{\pi^{3/2} a^{2(s-1)} \over \Gamma(s)} \left[
-{1 \over 2} (am)^{5/2-s} \Gamma\left( s -{5 \over 2} \right)
-{1 \over 2} J_3^{(0)}(s-1,am)
+J_3^{(1)}(s-1,am) ,
\right]
\eeq
where
\beq
\begin{array}{lll}
\ds J_M^{(0)}(z, \alpha)&=&\ds \int_0^{\infty} dt \ t^{z-M/2-1}
e^{-t \alpha^2} \left[ \theta\left( \pi \over t \right) -1 \right]
, \\
\ds J_M^{(1)}(z, \alpha)&=&\ds \int_0^{\infty} dt \ t^{z-M/2-1}
e^{-t \alpha^2} {d\over dt}\theta\left( \pi \over t \right) .
\end{array}
\eeq
Although not exactly like \req{defIM},
these integrals may also be written as Dirichlet series involving
modified Bessel functions. Furthermore, for $M=3$ such Bessel
functions
are expressible by finite series, and it is then possible to
interchange
the summations finally arriving at finite sums of polylogarithm
functions.
In this way, we get
\beq
\begin{array}{lll}
\ds \zeta_{{\bf R}^2\times {\bf S}^3}(1)&\ds =2\pi^2&\ds\left[
-{1 \over 3}(am)^{3/2}
-(am)^3 \left(
{ {\rm Li}_2(e^{-2\pi am}) \over (2 \pi am)^2 }
+ { {\rm Li}_3(e^{-2\pi am}) \over (2 \pi am)^3 }
\right) \right. \\
&&\ds\left. +\pi^2 am {\rm Li}_{-2}(e^{-2\pi am})
\right] , \\
\ds {\zeta_{{\bf R}^2\times {\bf S}^3}}'(0)&\ds =4\pi^2&\ds\left[
{1 \over 15}(am)^{5/2}
-(am)^5 \left(
{ {\rm Li}_3(e^{-2\pi am}) \over (2 \pi am)^3 }
+3 { {\rm Li}_4(e^{-2\pi am}) \over (2 \pi am)^4 }
+3 { {\rm Li}_5(e^{-2\pi am}) \over (2 \pi am)^5 }
\right) \right. \\
&&\ds\left. +\pi^2 (am)^3 {\rm Li}_{-1}(e^{-2\pi am})
\right] .
\end{array}
\eeq
Since we are in $D=5$, $\lambda_c$ must be
the value of $\lambda$ satisfying the second gap equation for
$b=0, m=0$, \ie
\[
{1 \over \lambda_c}=G(x,x;0,g)=
\left.
{1 \over (2\pi)^2 a^3} \zeta_{{\bf R}^2\times {\bf S}^3}(1)
\right\vert_{m=0}
=-{1 \over 16 \pi^3 a^3}{\rm Li}_3(1)
=-{1 \over 16 \pi^3 a^3}\zeta_R(3) .
\]
Next, we replace ${1 \over \lambda}$ with this critical value
and solve numerically the second gap equation for $b=0$ only.
The critical value obtained for $am$ is
\[
(am)_c = 2.2689.
\]
Calculating ${\zeta_{{\bf R}^2\times {\bf S}^3}}'(0)$ at
$am=(am)_c$, we find the finite contribution to the free energy
density
at the critical point:
\[
{W \over N}= -{1 \over 2}\left( a \over 2 \pi \right)^2
{\zeta_{{\bf R}^2\times {\bf S}^3}}'(0) = -0.7773.
\]

\subsection{Application to ${\bf T}^M\times {\bf S}^P$ \label{ssTMSP}}

The zeta function in this space may be written
\beq
\ds\zeta_{{\bf T}^M\times {\bf S}^P}(s)
= \left( 2 \pi \over \rho \right)^{-2s} \delta^{-s}
\sum_{n_1, \dots , n_M} \sum_l d(P,l)
\left\{
{1 \over \delta}(n_1^2 + \dots + n_M^2)
+\delta\left[ \left( l +{P-1 \over 2} \right)^2 +a^2m^2 \right]
\right\}^{-s},
\label{zrho}
\eeq
where we are using the notation
\beq
\delta = {\rho \over 2 \pi a} ,
\eeq
and
$d(P,l)$
is the degeneracy of the $l$th $P$-dimensional spherical mode
\req{degsph}.
In the $m=0$ case, using \req{dAks} this is put in the way
\beq
\begin{array}{c}
\ds \zeta_{{\bf T}^M\times {\bf S}^P}(s)
= \left( 2 \pi \over \rho \right)^{-2s} \delta^{-s}
\sum_{k=0}^{k_{\mbox{\ssz max}}(P)} (-1)^k {\cal A}_k(P) \\
\ds\times\sum_{n_1, \dots , n_M} \sum_l
\left( l +{P-1 \over 2} \right)^{P-1-2k}
\left[
{1 \over \delta}(n_1^2 + \dots + n_M^2)
+\delta \left( l +{P-1 \over 2} \right)^2
\right]^{-s},
\end{array}
\eeq
For odd $P$, ${P-1 \over 2}$ is an integer and
simple properties under dual transformations
$\delta \to {1 \over \delta}$
in the sense of sect. 4 in the second ref. of \cite{Ca} may appear,
after making the replacements explained there.
It is quite clear that it $P$ es even,
there can be no such invariance ---in whatever sense.
However, it can be achieved
considering antiperiodic (instead of periodic) field solutions,
which lead to the change $n_i \to n_i+{1 \over 2}$,
and making some further alterations. In particular, for
$M=1$, $P=2$, $d(2,l)=2(l+1/2)$ and $ l+(P-1)/2 =l+1/2$. As a result,
one can then put
$$
\ds\zeta_{{\bf T}^1\times {\bf S}^2}(s) \propto
\sum_{n,l} \left( l+{1 \over 2} \right) \left[
{1 \over \delta}\left( n+{1 \over 2} \right)^2
+\delta \left( l+{1 \over 2} \right)^2
\right]^{-s} .
$$
Then, the ensuing free energy will be invariant if one cares to replace
the ordinary $\delta$-derivative with the `fractional-derivative'
from \cite{Ca}. In a  similarl way, by
adequately modifying the definition of the free energy
we may get the $\delta\to {1 \over \delta}$ invariance
in higher dimensions.

Combining the general forms of the previous calculations we find
\beq
\ds\zeta_{{\bf T}^M\times {\bf S}^P}(s)
={ \pi^{M/2} \left( 2 \pi \over \rho \right)^{-2s} \over \Gamma(s) }
\sum_{l=0}^{\infty} d(P,l)
\left[ I_M\left( s, { \rho \over 2\pi} m(P,l) \right)
+\left( { \rho \over 2\pi } m(P,l) \right)^{M-2s}
\Gamma\left( s-{M \over 2} \right)
\right] ,
\eeq
where
\beq
m(P,l)\equiv
\sqrt{ {1 \over a^2} \left( l+{P-1 \over 2} \right)^2 +m^2 } .
\eeq
After expanding
$I_M$ into a Dirichlet series of modified Bessel functions, we
realize
that for $P\to\infty$ this part may be neglected
because it is exponentially vanishing. Calling
$\zeta_{{\bf T}^M \times {\bf S}^P}^{(\infty)}(s)$ the rest,
one gets
\beq
\zeta_{{\bf T}^M \times {\bf S}^N}^{(\infty)}(s) \sim
{ 2 a^{2s} \left( \rho \over a \right)^M \over (4\pi)^{M/2}}
{\Gamma\left( s-{M \over 2} \right) \over \Gamma(s)}
\times \left\{ \begin{array}{rll}
\ds \Sigma_{\alpha}(p, 2s-M), &\mbox{for $P=2p+2$}, \\
\ds \Sigma_{\beta}(p, 2s-M), &\mbox{for $P=2p+1$}, \\
\end{array} \right.
\eeq
which is, up to a constant, the zeta function for
${{\bf R}^M \times {\bf S}^P}$ with $m=0$, whose properties have
already
been considered.

\section{Higher-derivative $O(N)$ non-linear sigma model in
${\bf R}^{D-M}\times{\bf T}^M$}

It is interesting to observe that the model \req{Ztheor} may be
easily
generalized to have higher-derivative terms. In order not to have
to study
higher-derivative conformally invariant operators, we limit
ourselves
to ${\bf R}^{D-M}\times{\bf T}^M$, which is relatively simple due
to
its conformal flatness. Then we may write
\beq
Z[g] = \int {\cal D}\phi {\cal D}\sigma \exp\left\{
-{1 \over 2 \lambda} \int d^Dx \sqrt{-g} \left[
\phi^i \Box^2 \phi^i + \sigma (\phi^{i \ 2}-1) \right] \right\} .
\label{HDmodel}
\eeq
Repeating all the steps in section 1, we get
\beq
W[g, \lambda]=
{N \over 2}\left[ \Tr\log ( \Box^2+\sigma )
-\int d^Dx \sqrt{g} {m^2 \over \lambda} \right] .
\eeq
The Green function and the $\Tr\log$ part of the above expression
are directly linked to the
associated zeta function, which admits the following power
expansion in
$\rho\sigma^{1/4}$:
\beq
\begin{array}{c}
\ds\zeta^{( \Box^2+\sigma )}_{ {\bf R}^{D-M}\times{\bf T}^M }(s)=
{\pi^{D-M} \over \Gamma(s)} \sigma^{ {D-M \over 4} -s}\left[
{\sqrt{\pi} \over 2^{D-M \over 2}}
{ \Gamma\left( s-{D-M \over 4} \right) \over
\Gamma\left( D-M+2 \over 4 \right) } \right. \\
\ds\left. + {\sqrt{\pi} \over 2^{2s-1} }
\sum_{k=0}^{\infty} {(-1)^k \over k!}
{ \Gamma\left( 2s+2k-{D-M \over 2} \right) \over
2^{2k} \Gamma\left( s+k+{1 \over 2} \right) }
Z_M\left( 2s+2k-{D-M \over 2} \right)
\left( \rho \sigma^{1/4} \over 2 \pi \right)^{4s+4k-D+M}
\right] ,
\end{array}
\eeq
where $Z_M$ is the usual Epstein zeta function.
This function has poles at $s={D \over 4}-k$, $k=0,1,2,\dots$,
with the exception (if they coincide) of
$s=0, -1, -2, \dots$, where it is finite.

Mathematically speaking, this object is harder to deal with than
an `ordinary' zeta function (\ie one for a second-order operator).
It can also be expressed by a series of hypergeometric functions:
\beq
\begin{array}{c}
\ds\zeta^{( \Box^2+\sigma )}_{ {\bf R}^{D-M}\times{\bf T}^M }(s)=
{\pi^{D-M} \over 2^{D-M \over 2} \Gamma(s)}
\left( 2 \pi \over \rho \right)^{D-M-4s}
{ \Gamma\left( s-{D-M \over 4} \right)
\Gamma\left( s-{D-M \over 4}+{1 \over 2} \right) \over
\Gamma\left( s+{1 \over 2} \right) } \\
\ds\times\sum_{ \vec n \in {\bf Z}^M }
{}_2F_1\left(
s-{D-M \over 4},s-{D-M \over 4}+{1 \over 2},s+{1\over 2};
- \left( \rho\sigma^{1/4} \over 2 \pi \right)^4 (\vec{n}^2)^{-2}
\right)
(\vec{n}^2)^{{D-M\over 2}-2s} ,
\end{array}
\eeq
or by the integral representations
\beq
\begin{array}{c}
\ds\zeta^{( \Box^2+\sigma )}_{ {\bf R}^{D-M}\times{\bf T}^M }(s)=
{\pi^{D+1 \over 2} \sigma^{{s \over 2}+{1 \over 4}}
\over 2^{3s-1/2} \Gamma(s)}
\left( 2 \pi \over \rho \right)^{D-M-6s-1} \\
\ds\times \left[
{\cal I}_{D,M}\left( s, {\rho\sigma^{1/4} \over 2 \pi} \right)
+\left( \rho\sigma^{1/4} \over 2 \pi \right)^{D-6s-1}
{ \Gamma\left( 4s+1-{D \over 2} \right) \over
2^{{D+1 \over 2}-3s} \Gamma\left( -2s+2+{D \over 2} \right) }
\right] ,
\end{array}
\eeq
where
\beq
{\cal I}_{D,M}( s, \alpha )=
\sum_{l=1}^M \pmatrix{ M \cr l } 2^l
\sum_{ \vec{n}\in ({\bf N}^*)^l }
\int_0^{\infty} dt \ t^{3s-{D+1 \over 2}} J_{s+1/2}(\alpha^2 t)
\ e^{ -{\pi \over t} |\vec{n}|_l^2 } .
\eeq
Here $J$ is the first species Bessel function.

A different strategy is to regard $\Box^2+\sigma$ as
$(\Box+i\sigma^{1/2})(\Box-i\sigma^{1/2})$. Then,
\beq
{1 \over 2}\Tr\log( \Box^2+\sigma )=
{1 \over 2}\left[
\Tr\log(\Box+i\sigma^{1/2})
+\Tr\log(\Box-i\sigma^{1/2})
\right]
=-\left( \rho\over 2\pi \right)^{D-M}
\rep{\zeta^{( \Box+i\sigma^{1/2})}_{{\bf R}^{D-M}\times{\bf
T}^M}}'(0).
\eeq
Taking advantage of the calculation in \refs{sectRT} for
$-\Box+m^2$
when $D=2d+3$, we arrive at the following expression for the finite
part of ${W \over N}( \rho\sigma^{1/4} )$:
\beq
\begin{array}{ll}
\ds-{1 \over (4\pi)^{d+3/2} }&\ds\left\{
2^{d+3}\sqrt{\pi}
\sum_{k=0}^{d+1} { (d+1+k)! \over k! (d+1-k)! 2^k}
( \rho \sigma^{1/4} )^{d+1-k}
\rep\left[ i^{d+1-k \over 2}
{\rm Li}_{d+2+k}(e^{-i^{1/2}\rho\sigma^{1/4}} ) \right] \right. \\
&\ds\left. +\Gamma\left( -d-{3 \over 2} \right)
( \rho \sigma^{1/4} )^{2d+3} \rep[ i^{d+3/2} ]
\right\} .
\end{array}
\label{WNReps}
\eeq
Observing that the values at the critical point have to coincide
with the extremals of $W/N$ as a function of $\rho\sigma^{1/4}$,
these points are found by numerical examination of \req{WNReps}.
\begin{center}
\begin{tabular}{r|c|r}
$d$&min of
$-{1 \over 2}\left( \rho \over 2\pi \right)^{2d+2}
{\zeta_{{\bf R}^{2d+2}\times {\bf T}^1}}'(0)$&
$(\rho \sigma^{1/4})$ \\ \hline
1&1.65&-0.1459 \\
2&2.36&-0.1088 \\
\end{tabular}
\end{center}
For $d=$0,3 and 4 no solution has appeared, while for
$d=5$ we get 3.03 and -0.3141.  Comparing with numerical estimates
for the model \req{Ztheor} on the same background, we see that the
properties of the higher-derivative model \req{HDmodel} are drastically
different.

\section{Conclusions}
In the present paper, using zeta-function regularization, we have
calculated the free vacuum energy (or Casimir energy) for the
$D$-dimensional $O(N)$ non-linear sigma model on some spaces
with constant curvature. Explict expressions for the free energy
have been obtained at the critical point (when the gap equations
have been used) and also in non-critical regime (when the gap
equations have not been used). For all those spaces with $D=3,4$,
the explict expressions for the free energy may be easily used
in studies of the quantum antiferromagnet \cite{NePe}.
Moreover, for $2 < D <4$ it is known \cite{BZ} that the
$O(N)$-model in ${\bf R}^D$ possesses an order-disorder phase
transition and hence its free energy may be very useful for
studying dual properties. In this respect, the generalization
to curved backgrounds is of interest.

In particular, we have shown (see eq. \req{thisva}) that the free energy
of the model in ${\bf R}^{2q+1} \times {\bf S}^{2p+2}$ vanishes.
This generalizes the corresponding result of ref. \cite{GRV}
in ${\bf R}^{1} \times {\bf S}^{2}$, and shows that in $D$ dimensions
the free energy at the critical point
for ${\bf R}^{2q+1} \times {\bf S}^{2p+2}$ is
the same as that for ${\bf R}^D$, in accordance with the conformal
equivalence between both manifolds.

{}From another viewpoint, for the development of connections of
two-dimensional conformal field theory (which is a very powerful
tool) with higher dimensional models, it is important
to study the modular properties of such theories \cite{Ca}.
In subsect \ref{ssTMSP} we have checked dual invariance
in Cardy's sense for $m=0$ and antiperiodic modes in
${\bf S}^1\times {\bf S}^2$.  Hence, the higher-dimensional
$O(N)$-model may serve as a very useful toy model for studying
dual symmetries, which have become quite popular in recent
studies of strings and supersymmetric QCD.

Finally let us note that the $D$-dimensional $O(N)$ non-linear sigma
model may be considered as a toy model of quantum field theory in
a Kaluza-Klein framework for studying the question of spontaneous
compactification, for example. As usual, Kaluza-Klein theories are
not renormalizable. However, in the model under discussion we may
still use the ${1 \over N}$-expansion in order to control somehow the
quantum corrections.

Observe also that in addition to the ${1 \over N}$-expansion one
can calculate the free energy as an expansion in inverse powers
of the dimension. Such calculations in frames
of the toy model under consideration may be useful in atomic
physics \cite{BMP}.

\appendix\section{Appendix: on the calculation of zeta functions}

\subsection{General considerations}

Let ${\cal M}$ be a spacetime of the type
${\bf R}^{D-M}\times{\cal M}^M$, $D > M$, with
the operator $-\Box+m^2+\xi R$ acting on the whole manifold. Since
$-\Box=-\Box_{{\bf R}^{D-M}}-\Box_{{\cal M}^M}$, the spectrun has
the
form $p^2+\lambda_n+m^2+\xi R$, $p \in {\bf R}^{D-M}$,
$\lambda_n \in {\rm Sp}\left( -\Box_{{\cal M}^N} \right) $.
Then, the global zeta function
on ${\cal M}={\bf R}^{D-M}\times{\cal M}^M$ is defined as
\beq
\zeta_{{\bf R}^{D-M}\times{\cal M}^M}(s)=
\int d^{D-M}p \sum_{n: \atop \lambda_n \in {\rm Sp}(-\Box_{{\cal
M}^M})}
( p^2+\lambda_n+m^2+\xi R )^{-s} .
\eeq
Such a definition is purely mathematical, \ie
the usual physical factors coming from state normalization are here
absent. Notice that each momentum integration is introducing a
further
mass dimension.
The arbitrary mass scale $\mu$, typically supplied in order to
render
the function dimensionless, is not present either. All these
elements
may be included at a later stage.

After doing the $p$-integrations,
this zeta function can be written in terms of the one for ${\cal
M}^N$
\beq
\zeta_{{\bf R}^{D-M}\times{\cal M}^M}(s)=
\pi^{D-M \over 2}
{ \Gamma\left( s-{D-M \over 2} \right) \over \Gamma(s) }
\zeta_{{\cal M}^M}\left( s-{D-M \over 2} \right) ,
\label{Globalzf}
\eeq
with
\beq
\zeta_{{\cal M}^M}(z)= \sum_{n: \atop \lambda_n \in {\rm Sp}}
( \lambda_n+m^2+\xi R )^{-z} ,
\eeq
where possible degeneracies must be accounted for into this sum.

First, we make the following hypothesis: only the $\Gamma$
function has singularities at $s-{D-M \over 2}$, $s=0,1$,
---and, obviously, these poles can be encountered for even $D-M$
only---
while
$\zeta_{{\cal M}^M}$ and its derivative are finite at these points.
Such an assumption is right when
${\cal M}^M={\bf T}^M$ for odd $D$, or ${\cal M}^M={\bf S}^M$ for
the
massles case.
Using the expansion of the $\Gamma$ functions, both around their
poles
and around regular points,
we obtain the generic results:
\begin{enumerate}
\item{Even $D-M=2q$}
\beq
\begin{array}{lll}
\ds\zeta_{{\bf R}^{2q}\times{\cal M}^M}(1+\varepsilon )&=&
\ds
\pi^{q}
{(-1)^{q-1} \over (q-1)!}
\left\{
\left[ {1 \over \varepsilon}+\psi(q)+\gamma \right]
\zeta_{{\cal M}^M}\left( -q+1 \right)
+{\zeta_{{\cal M}^M}}'\left( -q+1 \right)
\right\}
+O( \varepsilon ), \\
\ds{\zeta_{{\bf R}^{2q}\times{\cal M}^M}}'(0)&=&
\ds
\pi^{q}
{(-1)^{q} \over q!}
\left\{
\left[ \psi(q)+\gamma \right]
\zeta_{{\cal M}^M}\left( -q \right)
+{\zeta_{{\cal M}^M}}'\left( -q \right)
\right\} .
\end{array}
\label{Globalzf1d0even}
\eeq
\item{Odd $D-M=2q+1$}
\beq
\begin{array}{lll}
\ds\zeta_{{\bf R}^{2q+1}\times{\cal M}^M}(1)&=&
\ds
\pi^{q+{1 \over 2}}
\Gamma\left( -q+{1 \over 2} \right)
\zeta_{{\cal M}^M}\left( -q+{1 \over 2} \right), \\
\ds{\zeta_{{\bf R}^{2q+1}\times{\cal M}^M}}'(0)&=&
\ds
\pi^{q+{1 \over 2}}
\Gamma\left( -q-{1 \over 2} \right)
\zeta_{{\cal M}^M}\left( -q-{1 \over 2} \right)
\end{array}
\label{Globalzf1d0odd}
\eeq
\end{enumerate}
When, on the contrary, $\zeta_{{\cal M}^M}(z)$ has poles at
$z=s-{D-M \over 2}$, $s=0,1$, this function has to be
Laurent-expanded
around these singularities before taking the limits $s \to 0,1$,
and
a somewhat different calculation is required in every particular
case.

\subsection{Zeta function on ${\bf T}^M$}

We study the torus only. Once we have the zeta function on
${\bf T}^M$, we can find the one on
$\zeta_{{\bf R}^{D-M}\times{\cal M}^M}$ by application of
\req{Globalzf}. So, when looking at $\zeta_{{\bf T}^N}(z)$ we
will have in mind the arguments $z=s-{D-M \over 2}, s=0,1$.

For simplicity we take
all the radii equal and with value $\rho$. Then
\beq
\zeta_{{\bf T}^M}(z)=
\left(2 \pi \over \rho \right)^{-2z}
Z_M\left( z , {\rho m \over 2\pi} \right)
\eeq
where
\beq
Z_M(z,a)=\sum_{n_1, \dots, n_M=-\infty}^{\infty}
(n_1^2+ \dots +n_M^2+a^2)^{-z}
\eeq
is an Epstein zeta function with inhomogeneous term.
After Mellin-transforming we obtain,
\beq
\zeta_{{\bf T}^M}(z)=
\left( 2 \pi \over \rho \right)^{-2z}{1 \over \Gamma(z)}
\int_0^{\infty} dt \ t^{z-1} \
e^{-t \left( \rho m \over 2\pi \right)^2}
\theta^M\left( t \over \pi \right) .
\eeq
The Jacobi theta function $\theta(x)$
---given by \req{deftheta}---
has the property
\beq \theta(x)={1 \over \sqrt{x}}\theta\left( 1 \over x \right) .
\eeq
Taking advantage of this, we separate the part which diverges at
$t=0$
---\ie the $n=0$ contribution in the $\theta$ function--- and write
\beq
\begin{array}{lll}
\ds\zeta_{{\bf T}^M}(z)&=&\ds
\left( 2 \pi \over \rho \right)^{-2z} {\pi^{M/2} \over \Gamma(z)}
\left\{
\int_0^{\infty} dt \ t^{z-M/2-1} \
e^{-t \left( \rho m \over 2\pi \right)^2}
\left[ \theta^M\left( \pi \over t \right) -1 \right]
+ \left( \rho m \over 2\pi \right)^{M-2z}
\Gamma\left( z-{M \over 2} \right)
\right\} \\
&\equiv&\ds
\left( 2 \pi \over \rho \right)^{-2z} {\pi^{M/2} \over \Gamma(z)}
\left\{
I_M\left( z, {\rho m \over 2\pi} \right)
+ \left( \rho m \over 2\pi \right)^{M-2z}
\Gamma\left( z-{M \over 2} \right)
\right\} .
\end{array}
\eeq
Now the $t$-integral, that (in accordance with \req{defIM}) we have
called $I_M\left( z, {\rho m \over 2\pi} \right)$,
contains no small-$t$
singularity. This expression has the additional advantage that its
second term exhibits the poles of this function at
$z={M \over 2}, {M-1 \over 2}, \dots $
After
writing $\theta^M$ as a binomial expansion (in the $n=0$ term of
$\theta$ and the rest of the summatory), we use the integral
representation
\beq
\int_0^{\infty} dt \ t^{\nu-1} e^{-{a \over t}-bt}
=2\left( a \over b \right)^{\nu/2} K_{\nu}(2 \sqrt{ab})
\eeq
and end up with the Dirichlet series
\beq
I_M\left( z, {\rho m \over 2\pi} \right)=
2^{M/2-z} \left( \rho m \over 2\pi \right)^{M-2z}
\sum_{l=1}^M \left( \begin{array}{c} M \\ l \end{array} \right)
2^{l+1} \sum_{n_1, \dots, n_l=1}^{\infty}
{ K_{M/2-z}( \rho m \sqrt{ n_1^2+ \dots +n_l^2 } ) \over
( \rho m \sqrt{ n_1^2+ \dots +n_l^2 } )^{M/2-z} } .
\eeq
These results, combined with \req{Globalzf}, give rise to
\req{zIMGa},\req{IMDir}
for $\zeta_{{\bf R}^{D-M}\times{\bf T}^M}(s)$.

\subsection{Zeta function on ${\bf S}^M$}
The same method is employed to derive the expressions in
$\zeta_{{\bf R}^{D-M}\times{\cal M}^M}(s)$ from the ones for
${\bf S}^M$.
Taking into account the known spectrum of the D'Alembertian on
${\bf S}^M$, the Riemann curvature on this space and the conformal
coupling value of $\xi$, we readily obtain the eigenvalues of our
operator and construct its zeta function
\beq
\zeta_{{\bf S}^{M}}(z)=\sum_{l=0}^{\infty}
d(M,l)
\left[ {1 \over a^2} \left( l+{M-1 \over 2} \right)^2 + m^2
\right]^{-z} .
\eeq
with degeneracies (see \eg \cite{Vil})
\beq
d(M,l)={ (l+M-2)! \over l! (M-1)! } (2l+M-1 )
=\left( \begin{array}{c} l+M-2 \\ l \end{array} \right)
{l+{M-1 \over 2} \over {M-1 \over 2}} .
\label{degsph}
\eeq

Setting $m=0$ one obtains
\beq
\begin{array}{l}
\ds \zeta_{{\bf S}^{M}}(z)=
{ 2a^{2z} \over M-1 } \sum_{l=0}^{\infty}
\left( \begin{array}{c} l+M-2 \\ l \end{array} \right)
\left( l+{M-1 \over 2} \right)^{1-2z}= \\
\left\{ \begin{array}{rll}
\ds {2a^{2z} \over (2p+1)!}&\ds\sum_{k=0}^{p} (-1)^k \alpha_k(p-1)
\zeta_H\left(-1+2z-2p+2k, p+{1 \over 2} \right),&\mbox{for
$M=2p+2$}, \\
\ds {2a^{2z} \over (2p)!}&\ds\sum_{k=0}^{p-1} (-1)^k \beta_k(p-2)
\zeta_H\left(2z-2p+2k, p \right),&\mbox{for $M=2p+1$},
\end{array} \right.
\end{array}
\label{zSMi}
\eeq
with the $\alpha_k$ and $\beta_k$ coefficients as written in
\req{defalphakbetak}.
The Hurwitz functions in \req{zSMi} may be reexpressed with the
help of
the identities
\beq
\begin{array}{lll}
\ds\zeta_H\left(z, p+{1 \over 2} \right)&=&
\ds\zeta_H\left(z, {1 \over 2} \right)
-\sum_{n=0}^{p-1}\left( n+{1 \over 2} \right)^{-z}, \
\zeta_H\left( z, {1 \over 2} \right)=(2^z-1)\zeta_R(z), \\
\ds\zeta_H\left( z, p \right)&=&
\ds\zeta_R\left( z \right) -\sum_{n=1}^{p-1}n^{-z} .
\end{array}
\label{relzhzr}
\eeq
Afterwards, taking advantage of the properties
\beq
\begin{array}{llll}
\ds \sum_{k=0}^{p} (-1)^k \alpha_k(p-1)
\left( n+{1 \over 2} \right)^{2(p-k)}&=&0,&
\mbox{ when $0 \leq n \leq p-1$}, \\
\ds \sum_{k=0}^{p-1} (-1)^k \beta_k(p-2)
\left( n+1 \right)^{2(p-1-k)}&=&0,&
\mbox{ when $0 \leq n \leq p-2$},
\end{array}
\label{vanishSumalphabeta}
\eeq
we may put
\beq
\zeta_{{\bf S}^{M}}(z)=
\left\{ \begin{array}{ll}
\ds 2a^{2z} \Sigma_{\alpha}(p,2z), &\mbox{for $M=2p+2$}, \\
\ds 2a^{2z} \Sigma_{\beta}(p,2z), &\mbox{for $M=2p+1$},
\end{array} \right.
\eeq
where the $\Sigma_{\alpha}(p,2z)$ and $\Sigma_{\beta}(p,2z)$
are the ones defined by \req{DefSigmaalphabeta}. Observing those
finite sums,
one locates the singularities of
$\zeta_{{\bf S}^{M}}(z)$, which come from the pole of
$\zeta_H(x,a)$
or $\zeta_R(x)$ at $x=1$:
\[
\left\{ \begin{array}{llll}
z&=&1, 2, \dots , p+1, &\mbox{for $N=2p+2$}, \\
z&=&\ds{3 \over 2},{5 \over 2},\dots ,p+{1 \over 2},&\mbox{for
$N=2p+1$}.
\end{array} \right.
\]
As a result, we realize that at the points we are interested in,
\ie, $z=s-{D-M \over 2}$, $s=0,1$, $\zeta_{{\bf S}^{M}}(z)$ is
finite
and we may therefore apply \req{Globalzf1d0even},
\req{Globalzf1d0odd},
thus getting the values of
$\zeta_{{\bf R}^{D-M}\times{\bf S}^M}(1)$,
${\zeta_{{\bf R}^{D-M}\times{\bf S}^M}}'(0)$ given in
\refs{sectRS}.

\vskip5ex
\noindent{\Large \bf Acknowledgements}

We would like to thank the referee for useful remarks.
A.R. is grateful to {\it Generalitat de Catalunya, Comissionat
per a Universitats i Recerca,} for a RED fellowship.

\end{document}